\documentstyle[prl,aps,epsf]{revtex} \def\narrowtext{} \tighten
\twocolumn
\input epsf.sty
\begin{document}
\draft

\title{Quasiparticles in the superconducting state of
$Bi_{2}Sr_{2}CaCu_{2}O_{8+\delta}$}
\author{
        A. Kaminski,$^{1}$
        J. Mesot,$^{2}$
        H. Fretwell,$^{1}$
        J. C. Campuzano,$^{1,2}$
        M. R. Norman,$^2$
        M. Randeria,$^3$
        H. Ding,$^{4}$
        T. Sato,$^5$
        T. Takahashi,$^5$
        T. Mochiku,$^6$
        K. Kadowaki,$^7$
        and H. Hoechst$^{8}$
       }
\address{
         (1) Department of Physics, University of Illinois at Chicago,
             Chicago, IL 60607\\
         (2) Materials Science Division, Argonne National Laboratory,
             Argonne, IL 60439 \\
         (3) Tata Institute of Fundamental Research, Mumbai 400005,
             India\\
         (4) Department of Physics, Boston College, Chestnut Hill, MA
             02467\\
         (5) Department of Physics, Tohoku University, 980-8578 Sendai,
             Japan\\
         (6) National Research Institute for Metals, Sengen, Tsukuba,
             Ibaraki 305, Japan\\
         (7) Institute of Materials Science, University of Tsukuba,
             Ibaraki 305, Japan\\
         (8) Synchrotron Radiation Center, Stoughton, WI, 53589\\
         }

\address{%
\begin{minipage}[t]{6.0in}
\begin{abstract}
Recent improvements in momentum resolution lead to qualitatively
new ARPES results on the spectra of $Bi_{2}Sr_{2}CaCu_{2}O_{8+\delta}$
(Bi2212) along the $(\pi,\pi)$ direction, where there is a node in the
superconducting gap.  We now see the intrinsic lineshape, which indicates
the presence of true quasiparticles at all Fermi momenta in the
superconducting state, and lack thereof in the normal state. The region
of momentum space probed here is relevant for charge transport,
motivating a comparison of our results to conductivity measurements
by infrared reflectivity.
\typeout{polish abstract}
\end{abstract}
\pacs{71.25.Hc, 74.25.Jb, 74.72.Hs, 79.60.Bm}
\end{minipage}}

\maketitle
\narrowtext

Landau's concept of the Fermi liquid\cite{NOZIERES} underlies much of
our present theoretical understanding of electron dynamics in crystalline
solids. Landau was able to demonstrate that, even though the electrons
interact strongly with one another, one can still describe the low
temperature properties of metals in terms of ``quasiparticle"
excitations,
which are bare electrons dressed by the medium in which they move. But we
now have materials, such as the high temperature superconductors (HTSCs),
and other low dimensional systems, where it is becoming increasingly
difficult to reconcile experimental results with the expectations of
Fermi
liquid theory \cite{ANDERSON,VARMA}.

Clearly, if the concept of quasiparticles is to
be useful, they must live long enough to be considered as
independent entities. In fact, in a Fermi liquid, the quasiparticles
at the Fermi momentum, $k_F$, have (at zero temperature) an infinite
lifetime at zero excitation energy - the Fermi energy, $E_F$ -  with
their lifetime decreasing quadratically with excitation
energy\cite{NOZIERES}. If one were to measure the spectral function
of the electrons at $k_{F}$, one would observe a broad feature,
corresponding to the incoherent part of the electron, with a sharp
peak at $E_F$ with spectral weight $z$, the quasiparticle component.
It is now well established by angle resolved photoemission spectroscopy
(ARPES) measurements that, despite the existence of a Fermi surface
in momentum space, there are no quasiparticles in the normal state of
optimally doped or slightly overdoped HTSCs near the $(\pi,0)$ point
of the Brillouin zone (see inset of Fig.~1) \cite{REVIEW1,REVIEW2}.
Below $T_c$, the superconducting gap is maximal here (at the anti-node
point $A$) and sharp quasiparticle peaks are observed \cite{NK}. This
statement can be made since the energy dispersion of the electronic
states in this region of the zone is very weak, and therefore the
measured spectra are not artificially broadened by the finite
acceptance angle (momentum window) of the detector.

Along the zone diagonal, however, the intrinsic lineshape is unknown,
both
in the normal and superconducting states, because the spectra near
point N (at the Fermi surface along the diagonal; see inset of Fig.~1)
are significantly broadened by the momentum window given the highly
dispersive nature (1.6 eV$\AA$) of the states in this region.
This observation is reinforced by the lack of any temperature dependence
of the spectra at point N in sharp contrast to what is observed at point
A. In this work, a large improvement in experimental momentum resolution
allows us to determine the intrinsic lineshape at $N$ for the first time.
The significance of these observations stems from the fact that this
region of the zone dominates many of the bulk properties of cuprate
superconductors, and it is very important to know if and when
quasiparticles exist. Above $T_c$, the charge and thermal transport
are dominated by these states because of their 
rapid dispersion (large Fermi velocity). 
Below $T_c$, the superconducting energy gap vanishes
at the nodal point $N$, and low energy excitations in its neighborhood
dominate the $T$-dependence of various properties, e.g., the
superfluid density\cite{HARDY} and thermal Hall conductivity\cite{ONG}.
Moreover, there has been a gamut of opinions concerning the nature
of the electronic states near the node, ranging from the cold spot model
\cite{PINES} where quasiparticles are assumed to be present even above
$T_c$, and the stripes model\cite{STRIPES}, where quasiparticles do not
exist
even below $T_c$.

Measurements were carried out at the SRC, Wisconsin, on a
4m NIM undulator beamline (resolving power of
$10^{4}$ at $10^{12}$ photons/sec) as well as a PGM beamline. We used  a
Scienta SES 200 analyzer in angle resolved mode. The angular 
resolution and spacing of EDCs is 0.0097 $\AA^{-1}$ at 22eV as 
calibrated by measuring the superlattice wavevector. The energy 
resolution was 16 meV (FWHM). Samples were mounted with the
$\Gamma X$ direction parallel to the photon polarization,
except for Fig.~2 right panel and Fig.~1 left panel, where samples
were aligned along $\Gamma \bar{M}$.
ARPES gives direct information about the momentum and energy
dependence of the lifetime of electrons.
In quasi-2D materials, the ARPES signal is given by\cite{NK}
$I(\omega)=I_{\bf k}f(\omega)A({\bf k},\omega)$, convolved with the
energy resolution and momentum aperture of the detector.
Here $I_{\bf k}$
is the dipole matrix element between initial and final states,
$f(\omega)$
the Fermi function, and $A({\bf k},\omega)$ the spectral function.

\begin{figure}
\epsfxsize=3.3in
\epsfbox{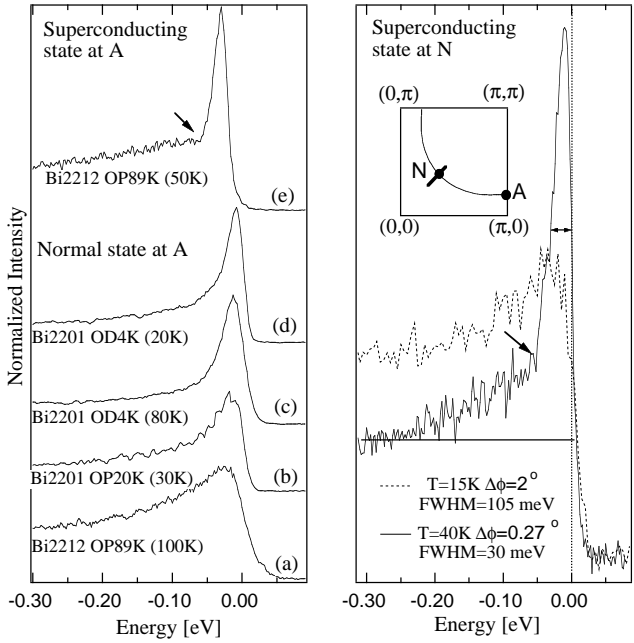}
\vspace{0.3cm}
\caption{
Left panel: ARPES spectra at
point A of the zone (inset), with spectra
labeled by doping (OD for overdoped, OP for optimal doped)
and onset $T_{c}$. The temperature is given in parenthesis. Right panel: 
ARPES spectra of
Bi2212 ($T_c$=89K) in the superconducting state at point N for 
two values of the resolution. We define FWHM of the peak
by the horizontal arrow
assuming that peak height is measured from a horizontal line drawn
at the background level.  Note the break in the high resolution spectrum
marked by an arrow.}
\label{fig1}
\end{figure}

Curve (a) in the left panel
of Fig.~1 shows an ARPES spectrum in the normal state at 100K for an
optimally doped Bi2212 sample with $T_{c}=89$ K.
In principle, at the Fermi surface ($k=k_F$), $A({\bf k},\omega)$
should have a peak centered at zero binding energy. Since ARPES
only measures the occupied part of $A$, the spectral peak
is cut off by the Fermi function, and thus its maximum is displaced to
higher energy. Therefore an estimate of its full width-half maximum (FWHM)
can be obtained by doubling the measured one, yielding
a value of $\sim$200 meV.
Although quasiparticles are only expected to appear at low
temperatures, we note this width is of order 2000K, well over
an order of magnitude larger than the temperature, indicating that
thermal broadening alone cannot be responsible for the large peak width.

We can confirm that large widths are intrinsic to {\it optimally} doped
cuprates by examining the spectral function of the single CuO layer
compound Bi2201, which has a lower $T_c$, and therefore a normal state
accessible at a lower temperature. In curve (b) of Fig.~1 we plot ARPES
data for the slightly overdoped Bi2201 compound
$Bi_{1.6}Pb_{0.4}Sr_{2}CuO_{6}$ ($T_{c}=20K$). Even though the normal
state data are now taken at 30K, the spectral width has only narrowed
to a FWHM of 100 meV, and so does not exhibit a quasiparticle peak.
We therefore conclude that quasiparticles do not appear in the normal
state of optimally doped compounds at point $A$, even at low temperature.
As there is some evidence from transport that more Fermi liquid like
behavior develops for heavily overdoped materials, the question arises
whether ARPES sees evidence for normal state quasiparticles in that case.
Curve (d) of Fig.~1 shows a spectrum for a highly overdoped Bi2201
sample ($T_c$=4K) at 20K. Indeed, the spectral
peak is much narrower than the optimally doped sample, consistent
with more Fermi liquid like behavior.
It can be seen that the broadening of the peak in the
optimally doped sample is not of thermal origin by comparing it to the
overdoped sample at a higher temperature, shown in curve (c) of Fig.~1.
The width of the peak of the overdoped sample at 80K is in fact
narrower than that of the optimally doped sample at 30K.

Although the normal state of optimally doped cuprates does not exhibit
them, quasiparticles do appear in the superconducting state at point $A$. 
Curve (e) in Fig.~1 shows a spectrum in the 
superconducting state, where a near resolution limited peak appears.  In
addition, a break clearly separates the coherent quasiparticle part
of
the spectral function from the incoherent part, as indicated by the
arrow \cite{FOOTNOTE}.
A significant point is that in the vicinity of the $A$ point, the break
in the spectra appears exactly at $T_{c}$ \cite{PHENOMENOLOGY}. This is
thought to be due to the onset of superconductivity, which leads to a
reduction of the scattering rate of electrons over an energy range of
order 2-3 times the maximum superconducting gap \cite{MODE}, thus
allowing
quasiparticles to exist.

It is now well established that the HTSCs are $d$-wave 
superconductors with points N characterized by nodes in the energy gap
\cite{REVIEW1,REVIEW2}. We now ask the question whether quasiparticles
exist at these particular points which exhibit gapless
excitations. So far, it has not been possible to
address this question because of resolution issues. For example, the
dashed curve in the right panel of Fig.~1 shows the spectrum
obtained at $N$ with the momentum resolution of all previously published
work, deep in the
superconducting state (T=15K). Although this peak has been called a
``quasiparticle peak" in the literature, there is no direct evidence for
this in the raw data. The spectrum is as broad as the normal state
spectrum of curve (a) in Fig.~1, but for a different reason. This is the
direction in momentum space of largest dispersion, and the finite
momentum window $\delta k$ of the analyzer broadens the peaks as
$\delta E=\hbar v_{F}\delta k$, where $v_F$ is the Fermi velocity. From
the observed dispersion, this momentum broadening is
 of the order of 100 meV.

\begin{figure}
\epsfxsize=3.3in
\epsfbox{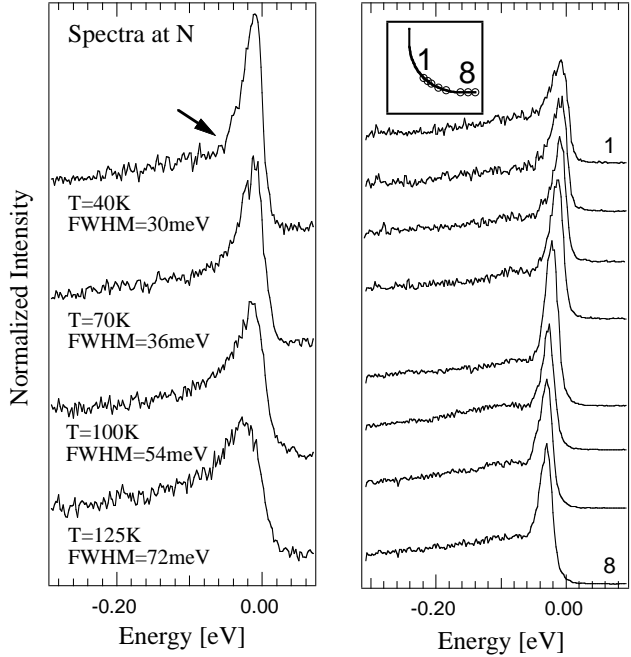}
\vspace{0.3cm}
\caption{
Left panel: Temperature dependence of the spectra at point N (inset
of Fig.~1) for a Bi2212 sample ($T_c$=89K). A break in the lineshape of
the spectra at 40K is indicated by an arrow. Right panel: Comparison
of the superconducting state spectra (T=40K) at various points along
the Fermi surface.}
\label{fig2}
\end{figure}

Now, with a 32-fold improvement in momentum resolution from the Scienta
detector (8 times along
the direction normal to the Fermi surface, and 4 times in the transverse
direction), we show in the solid curve of the right panel of Fig.~1 the
qualitatively new result that there are true quasiparticles at the nodes
of the d-wave superconducting state\cite{NODE}.
Although these data were obtained in the superconducting state, because
they are at the node of the $d$-wave state, they are in fact gapless.
A break separating the coherent from the incoherent part of the spectral
function is visible in the lineshape as indicated by an arrow. But the
converse is also true: there are no signs of quasiparticles in the normal
state. Fig. 2 (left panel) shows the temperature dependence of the
lineshape at point $N$.
We note that there is a distinct change in lineshape due to the
appearance of quasiparticles in the optimally doped sample.
In the normal state, the trailing edge of the spectral peak smoothly
evolves into an incoherent tail going to high binding energy.  As the
temperature is lowered below $T_c$, a break develops which
separates the coherent quasiparticle peak from the incoherent tail.
It is important to note that quasiparticles exist all along the Fermi
surface (Fig. 2b). 

\begin{figure}
\epsfxsize=2.9in
\epsfbox{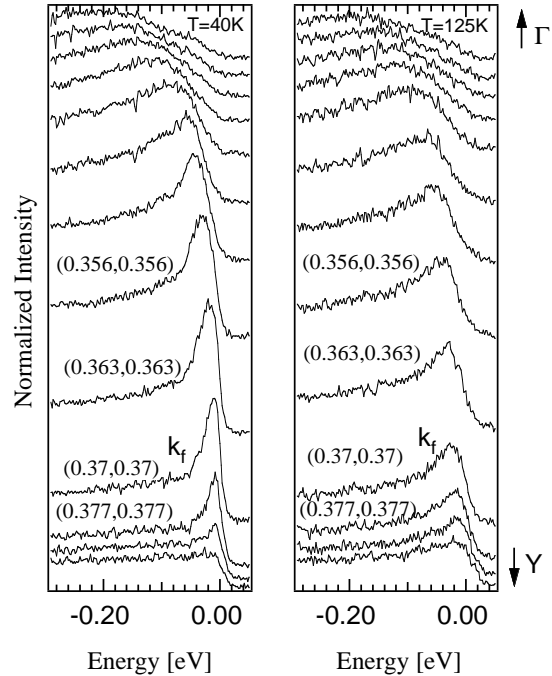}
\vspace{0.3cm}
\caption{
Momentum dependence of the spectra of Bi2212 ($T_c$=89K)
along the $(\pi,\pi)$ direction -indicated by the dark line segment in
the
inset of Fig.~1- in the normal (125K) and superconducting (40K) states.
Data were
taken at intervals of 0.0097 $\AA^{-1}$.  The spectrum
corresponding to the Fermi momentum is labelled $k_f$ and selected
spectra are labeled in $\pi/a$ units.}
\label{fig3}
\end{figure}

In Fig.~3, we plot the momentum dependence of the
lineshape along the zone diagonal (at $45^\circ$ to the Cu-O bond
direction) in the normal (125K) and superconducting (40K) states for an
optimally doped 89K sample. We note the sharpening of the spectral peak in the
normal state as $k_F$ is approached, as observed before \cite{OLSON}.
In the superconducting state, we see the new result of a coherent peak
along this direction, which only exists in a narrow momentum interval
about $k_F$.
More quantitative information can
be obtained by plotting the FWHM of the spectral peak from the raw data
as a function of the binding energy of the peak, as shown in
Fig.~4. Above $T_c$, the FWHM is approximately linear in
binding
energy (we define the FWHM relative to the horizontal line in the right
panel
of Fig. 1), with a slope of 0.5.  An extrapolation of the linear part to
zero binding energy results in an offset of 80 meV, about an order of
magnitude larger than the temperature.
Below $T_c$, the FWHM is the same as that in the
normal state for binding energies above an energy of about 2-3 times the
maximum superconducting gap ($\Delta_{max}$=40 meV), but decreases faster
than this below,
as expected for electron-electron scattering\cite{3DELTA}.
The residual offset at zero binding energy is
a combination of momentum and energy resolution, and a contribution
from the incoherent tail of the spectra (the FWHM of the coherent peak
is 30 meV, compared to a resolution estimate of 22 meV).

\begin{figure}
\epsfxsize=3.3in
\epsfbox{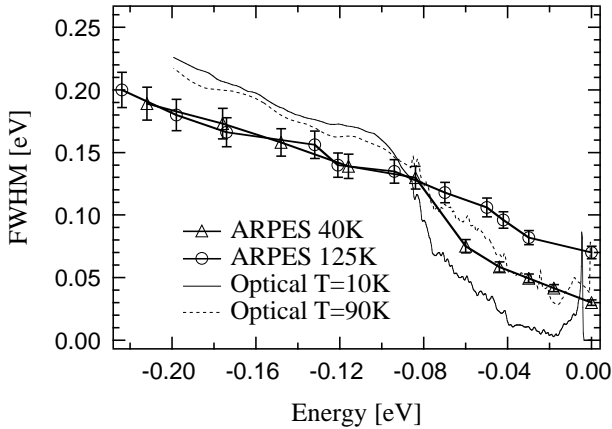}
\vspace{0.3cm}
\caption{
Full width-half maximum of the spectral peaks in Fig.~3 versus the
binding energy of the spectral peak (symbols), and the carrier
scattering rate versus energy for Bi2212 ($T_c$=90K) obtained from
infrared reflectivity measurements (solid and dashed lines)
\protect\cite{TIMUSK}. The FWHM is defined by the horizontal arrow
in the right panel of Fig.~1.}
\label{fig4}
\end{figure}

A detailed comparison of ARPES data to optical conductivity data would
require fitting the spectra to a model self-energy, using this self-energy
to construct the transport scattering rate, $1/\tau$, and then averaging
$1/\tau$ over the Brillouin zone with velocity weighting
factors\cite{TIMUSK}.  Rather than go through such a complicated
procedure, we elect to directly compare the $1/\tau$ from optical
conductivity in Bi2212 to the FWHM versus binding energy (discussed
above), which should be a rough measure of $1/\tau$ versus energy.
This is also shown in Fig.~4, where a good match is seen between these
two
quantities (at low energies, there is a deviation due to the ARPES
resolution).
We observe that the break in the spectrum near $k_F$ 
separating the coherent
peak from the incoherent tail occurs at the same energy, indicating
a drop in the imaginary part of the self-energy at the same frequency
that optical data see a drop in $1/\tau$.
Finally, we note that the linear energy variation seen by the optical and
APRES data is analogous to the linear temperature dependence of the resistivity,
ubiquitous in the cuprate superconductors, and is in support of a marginal Fermi
liquid
 phenomenology\cite{VARMA}.

In conclusion, we report the first observation of the intrinsic lineshape
along the zone diagonal, from which we deduce a quasiparticle peak
below $T_c$ and the absence of such a peak in the normal state.
The non-existence of quasiparticles in the normal state
appears to be a general property of cuprate superconductors.
Therefore, the formation of quasiparticles must be related to the
presence of a {\it superconducting} state. A proper description of
high temperature superconductivity will only arise after this peculiar
phenomenon is understood.

Note in proof:  After the completion of this work, we became aware of
similar work by T. Valla {\it et al.}, Science {\bf 285}, 2110 (1999),
with somewhat different conclusions.

We thank T. Timusk for providing the optical data. This work was supported 
by the National Science Foundation DMR 9624048, 
and DMR 91-20000 through the Science and Technology Center for
Superconductivity, the U. S. Dept. of Energy, Basic Energy Sciences,
under contract W-31-109-ENG-38, the CREST of JST, and the Ministry of
Education, Science, and Culture of Japan. The Synchrotron Radiation
Center
is supported by NSF DMR 9212658. JM is supported by the Swiss National
Science Foundation, and MR in part by the Indian DST through the 
Swarnajayanti scheme.


\begin{references}
\bibitem{FOOTNOTE}
The separation of the spectral function into a quasiparticle piece and 
an incoherent part can be justified from the breakup of the Green's function 
into corresponding singular and regular parts 
(Ref.~\cite{NOZIERES}).
\bibitem{NODE}
Note the node itself is a single point,
so makes a contribution of measure zero to the spectrum.
\end{references}
\end{document}